# OVERVIEW OF MC CDMA PAPR REDUCTION TECHNIQUES

B.Sarala[1] and D.S.Venkateswarulu[2] B.N.Bhandari[3]

[1]Department of ECE, M V S R Engineering College, Hyderabad
b.sarala@rediffmail.com
[2]Department of ECE, Progressive Engineering College, Cheekati Mamidi, HMDA, Hyderabad.
dsv4940@gmail.com
[3]Department of ECE, JNTU, Hyderabad, India.
[3]bnb@ieee.org

**ABSTRACT**

*High Peak to Average Power Ratio (PAPR) of the transmitted signal is a critical problem in multicarrier modulation systems (MCM) such as Orthogonal Frequency Division Multiplexing (OFDM), and Multi-Carrier Code Division Multiple Access (MC CDMA) systems, due to large number of subcarriers. High PAPR leads to reduced resolution, and battery life. It also deteriorates system performance. This paper focuses on review of different PAPR reduction techniques with attendant technical issues as well as criteria for selection of PAPR reduction technique. To reduce PAPR the constraints are low power consumption, and low Bit Error Rate (BER). Spectral bandwidth is improved by better spectral characteristics, and low complexity/cost.*

## 1. INTRODUCTION

MC CDMA system is the combination of OFDM and Code Division Multiple Access (CDMA) and reaps the benefits of both the techniques. In MC CDMA, data symbols consisting of modulated bits are spread by spreading codes and then mapped into subcarriers of a MC CDMA modem data symbol which is spread across frequency domain [1]. It is a very attractive technique for high speed data transmission over multipath fading channels. It improves security, minimizes multipath effects, such as Inter symbol Interference (ISI) and Inter Carrier Interference (ICI) [2]. MC CDMA and OFDM systems are widely used in the existing 3rd and 4th generation wireless networks. They are also good candidates for the future generation networks for broadband and personal communications. MC CDMA system is widely used for Long Term Evaluation (LTE), WiMAX, and Digital TV transmission [3]. The technical issues in MC CDMA are time dispersion, synchronization, Doppler spreading, frequency and phase offset subcarrier selection and high PAPR. Multicarrier modulation system applications are Digital data Transmission over the Telephone system, Digital audio broadcasting, Digital Television, and Wireless Local Area Networks [4]. However, MC CDMA systems have the inherent problem of a high PAPR, which causes serious performance degradation in the transmitted signal.

When the signal passes through the High Power Amplifier (HPA), the high PAPR causes the peaks to enter into saturation region resulting in in-band radiation (IBR) and out-of- band radiation (OBR). The IBR degrades the performance by increasing BER, but the OBR results in Adjacent Channel Interference (ACI). The PAPR brings disadvantages like the design complexity of Analog to Digital Converter (ADC) and Digital to Analog Converter (DAC).It also reduces power efficiency, increase BER and consumption of power. Use of HPAs result in increased cost, reduced battery life, increased co-channel interference and Inter Symbol Interference (ISI).Due to non-linear distortion in HPA; it further results in a loss of subcarrier orthogonality and spectral regrowth. Thus, if we reduce PAPR, we shall obtain reduced complexity of ADC and DAC, improved Signal to Noise Ratio (SNR) and BER [5, 6], enhancement of bandwidth and battery life with low power consumption. There has been a lot of research work done on PAPR reduction techniques in MCM systems. Several techniques are proposed for PAPR reduction [7].

The first one is the signal distortion technique, which introduces distortion to signals and causes degradation in the performance including clipping, clipping and filtering, windowing, peak cancelling, pre-distortion or companding. Signal pre-distortion techniques based on companding to reduce the PAPR have been proposed by several authors using different companding techniques such as μ-Law [8], exponential [9], modified exponential [10] and linear companding [11]. In companding technique, compression in transmitter and expansion in receiver has been proposed by Wang et al [12]. Clipping is simple and effective and causes IBR and increased BER. The companding transforms' are better than clipping as they reduce distortion significantly. Yuan Jiang [7] proposed an algorithm that uses the special airy function and is able to provide an improved BER and minimized OBI in order to reduce PAPR effectively.

Different coding techniques are proposed for signal scrambling which can be classified into: schemes with explicit side information including linear block codes, and multiple signal representations such as Selective Mapping (SLM) and Tone Reservation (TR), Partial Transmit Sequence (PTS), and interleaving schemes [13]. These schemes give an ease of modification, at increased overhead, search complexity, and data loss. The various schemes proposed without side information include block coding, Hadamard transform method, dummy sequence insertion method, Golay complementary codes, Reed Muller codes etc. Signal scrambling technique may not be affecting the system performance but it has an overhead of increased complexity and needs to perform exhaustive search to find best codes and to store large lookup tables for encoding and decoding. It does not support error correction [8, 9, and 10].Tone Reservation (TR) offers a distortion-less technique that effectively reduces the PAPR without side information, but BER increases. Coding can also be used to reduce the PAPR by selecting suitable code words that minimize the PAPR of the transmitted signal. Error control selective mapping is effective and there is no need for side information but, complexity is increased. Other schemes are modified signal constellation and Pilot tone method. The method proposed using separation of complex baseband signal for all modulations improves performance and results in less complexity [14]. Vijayarangan et.al [15] proposed pulse shaping Raised Cosine (RC) and Root Raised Cosine pulse waveforms for PAPR reduction without side information at the receiver, resulting no IBR and OBR. Pulse shaping schemes are very efficient and flexible, with less implementation complexity.

Seung Han et.al [13] describe comprehensively about distortion and distortion-less signal scrambling, active constellation PAPR reduction techniques for OFDM and MIMO-OFDM. However it does not include precoding Transform and spreading techniques. This paper focuses on some of the important PAPR reduction techniques viz., clipping and filtering, companding, PTS and SLM, precoded transform techniques such as Discrete Cosine Transform (DCT),

Discrete Fourier Transform (DFT), Discrete Hartley Transform (DHT) and Discrete Wavelet Transform (DWT) and spreading sequences. This paper also discusses the selection criteria for spreading sequence and PAPR reduction techniques.

The rest of the paper is organized as follows: Section 2 describes PAPR of MC CDMA signal and section 3 describes Complementary Cumulative Distribution Function (CCDF) of the PAPR. PAPR models are discussed in section 4 with the help of suitable algorithms and block diagrams. In section 5 spreading sequence selection criteria and Section 6 criteria for selecting PAPR reduction techniques are discussed. Conclusions are presented in section 7.

## 2. PAPR OF A MC CDMA SIGNAL

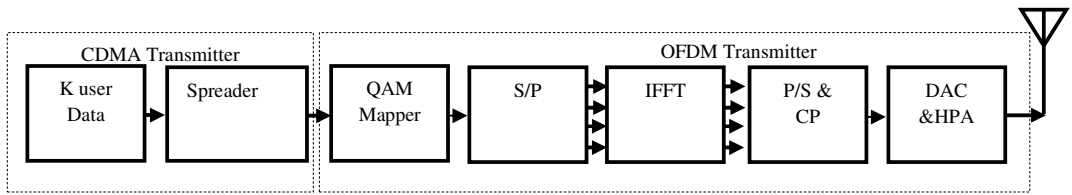

Figure1 MC CDMA Transmitter

Figure 1 shows the MC CDMA transmitter for down-link of the $k^{th}$ user. It transmits data symbol of a K user simultaneously on several narrow band sub channels. In MC CDMA K user data symbols are spread, modulated and mapped in the frequency domain using Inverse Fast Fourier Transform (IFFT). At the spreader the k user data is spread in the time domain by $k^{th}$ user's spreading sequence, then applied to Quadrature Amplitude Modulation (QAM) mapper followed by serial to parallel (S/P) converter and then applied to IFFT in the frequency domain. MC CDMA uses IFFT to divide the bandwidth into orthogonal overlapping Nc subcarriers each of the Nc subcarriers are modulated by a single chip. The data is converted back into serial data before cyclic prefix or guard interval is inserted to combat ISI. Finally the signal is fed to DAC followed by HPA, and up converter for transmission.

Generally, the PAPR of the MC CDMA signal p (t) is defined as the ratio between maximum instaneous power and its average power during the MC CDMA signal [16].

$$PAPR = \frac{\max[|p(t)|^2]}{E[|p(t)|^2]} \qquad (1)$$

Where E [.] denotes expectation and complementary cumulative distribution function for MC CDMA signal can be written as CCDF = probability (PAPR> P0), where P0 is the Threshold [12].

PAPR of MC CDMA signal is mathematically defined as

$$PAPR = 10 \, log_{10} \frac{\max[|p(t)|^2]}{\frac{1}{T} \int_0^T |p(t)|^2 dt} \qquad dB \qquad (2)$$

It is easy to manipulate the above equation by decreasing the numerator max [|p (t)|²] or increasing the denominator E[|p(t)|²] or both.

## 3. CCDF OF THE PAPR

The Complementary Cumulative distribution (CCDF) of the PAPR is one of the most regularly used parameter for measuring the performance of the MC CDMA system. The CCDF of the PAPR represents the probability that the PAPR of a data block exceeds a given threshold [17, 18]. The CCDF of the PAPR of a data block with Nyquist rate sampling is given as

$$P(PAPR>Z) = 1-P(PAPR <=Z)$$

$$= 1-F(Z) N_c$$

$$= 1-(1-\exp(-Z)) N_c.$$

This expression is not accurate for a small number of subcarriers.

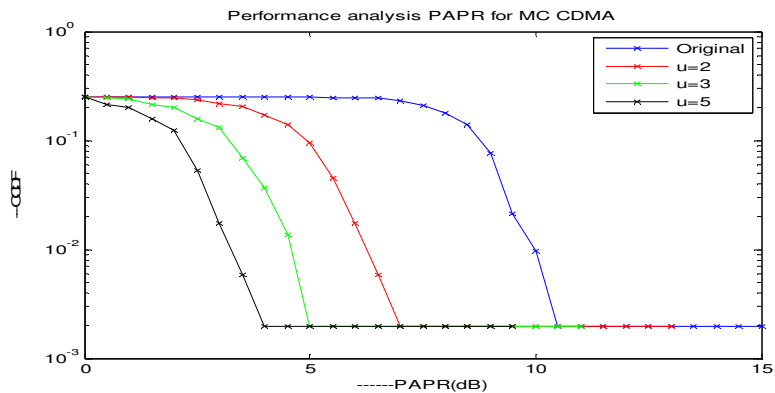

Figure 2 CCDF of PAPR of MC CDMA signal with companding with 64 subcarriers, IFFT 128,512 symbols and QPSK modulation.

Figure 2 shows the CCDF performance of MC CDMA original system [16], MC CDMA with companding. The figure shows that when CCDF less than $10^{-3}$ in MC CDMA with the companding method, the PAPR is reduced by 3.5dB, 5.5 dB, and 6.5 dB for values of μ is 2, 3, and 5 respectively when compared with the original MC CDMA system.

## 4. PAPR MODELS

### 4.1 Clipping and Filtering

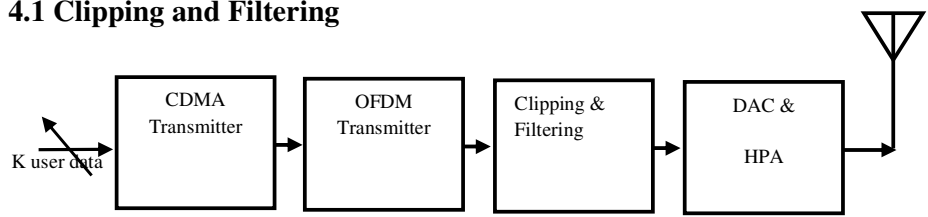

Figure 3 MC CDMA with clipping & Filtering

Figure 3 shows MC CDMA transmitter is combination of CDMA and OFDM transmitters [16, 35]. The MC CDMA transmitted signal is fed to Clipping followed by filtering then the signal is processed through DAC and HPA. Clipping is simple and effective by selecting optimum clipping ratio to remove the high amplitude peaks. But, it degrades system performance by introducing IBR and OBR, which can be reduced by filtering. This technique results in peak regrowth and distortion of the transmitted signal that can be reduced by repeated clipping and filtering. Using filtering technique improves IBR without degradation of BER. Using these technique additional blocks are used for clipping and frequency domain filtering at the transmitter, whereas receiver is unchanged. Transmitter complexity is increased by adding Fast Fourier Transform (FFT), IFFT filtering and computationally expensive. This technique is simple compared with companding, and PTS, SLM [19, 29].

## 4.2 Companding

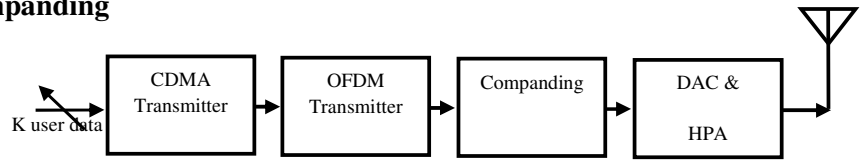

Figure 4 MC CDMA with Companding

Figure 4 shows MC CDMA with companding technique. The companding transformation is applied at the transmitter after IFFT block in order to attenuate the high peaks and amplify low amplitude of the MC CDMA signal, thus decreasing the PAPR. At the receiver, the de-companding process is applied by using the inverse companding function prior to FFT block in order to recover the original signal. The transmitted signal power is amplified by using HPA. However, fluctuations of signal amplitudes require expensive HPAs with very good linearity, large PAPR requires ADC and DAC with large dynamic range. Companding techniques are effective and simple for reduction of PAPR in MC CDMA. Companding technique describes compression in transmitter and expansion in receiver. μ-law companding technique, which enlarges only small signals so that average power increases, cause side lobe generation. Exponential companding scheme maintains constant average power and causes less spectrum side lobes. Exponential companding scheme offers better PAPR reduction, BER, phase error performance than μ-law companding technique[20].Using companding technique reduces PAPR, better Power spectral density, low implementation complexity and no constraints on modulation format and subcarrier size [1, 6, and 21, 22]. It has less complexity than SLM and PTS schemes. Companding technique increases the average received power and OBR. Transmitter and receiver need a compander and expander.

## 4.3 The Partial Transmit sequence

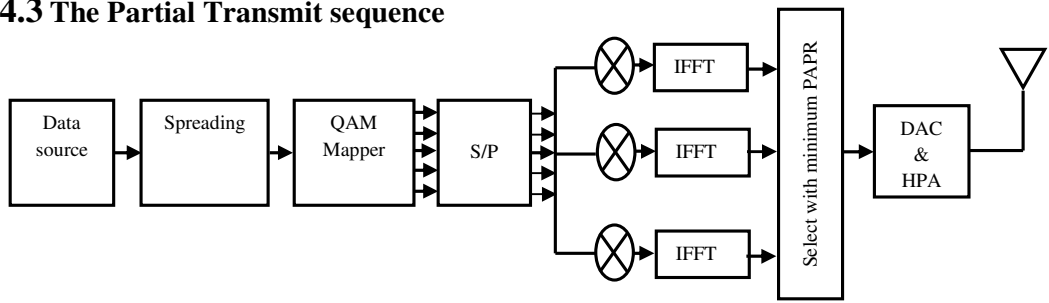

Figure 5 MC CDMA with PTS

The main idea is based on the coordination of proper phase rotation of signal parts to minimize the peak power of a MC CDMA transmitted signal. Jan hou et.al [22] proposed a novel PTS scheme, which simplifies the computation for each candidate signal without reducing the number of candidate signals. It can reduce computational complexity for the same PAPR over conventional PTS scheme (complexity increases with the increased number of candidate signals).Figure 5 shows MC CDMA with PTS scheme [13, 16, 22]. The input data sequence is divided into a number of disjoint sub blocks, which are weighted by phase factors to create a set of candidate signals, finally, the candidate with the lower PAPR is chosen for transmission. While using exhaustive search method the selection of optimum set of phase factors that minimize PAPR, increases search complexity exponentially as the number of sub carriers increases. To reduce search complexity, many modified techniques are proposed on reducing the number of candidate signals with less complexity.PTS sub-blocking schemes are adjacent, interleaved, and Pseudo random.PTS scheme works with arbitrary number of sub carriers and QAM mapper scheme. It uses Nc sub-carriers which are divided into Nc/2 sub blocks, and requires Nc/2 IFFTs. Nc/2 -1 bit information is required at the receiver to recover the data. Using this scheme circuit complexity and cost are increased. Additional side information is required at the receiver to recover the data there by reduces data rate. Computational complexity increases by increasing number of blocks, followed by data loss for using side information [22, 23].

### 4.4 The Selective Mapping

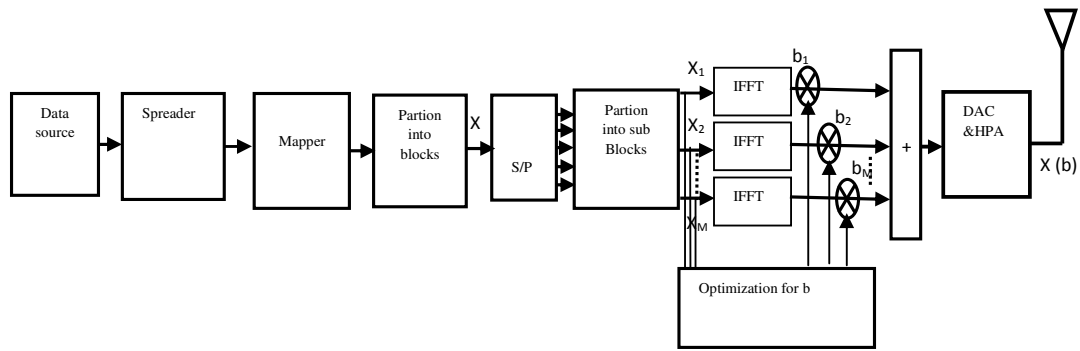

Figure 6 MC CDMA with SLM

The selected mapping technique consists of adding a set of phases in the transmitted signal in order to avoid phase alignment. SLM technique is proposed by Baumel et.al [24].PAPR reduction of MC CDMA proposed by Sajjad et.al [3], reduces PAPR very much and system performance degrades by increasing PAPR control bits (optimum is required). Figure 6 shows MC CDMA with SLM scheme [13, 16], SLM scheme is a PAPR reduction technique which is based on phase rotation (where phase sequences are generated randomly from set) to generate number of candidate signals. SLM Multicarrier Modulation (MCM) system has a set of V different blocks, all representing the same information and selects minimum PAPR block for transmission. Each data block is multiplied with V different phase sequences, each of length N resulting V modified data blocks and include one un-modified data block. One with lowest PAPR is selected for transmission. The selected phase information is transmitted to the receiver as side information to recover the original data. SLM schemes applied for QAM mapper and the amount of PAPR reduction depends on design of phase sequences. SLM needs number of iterations to find optimum phase factor for reduction of the PAPR of the transmitted signal. Adaptive PTS has been proposed to reduce the number of iterations by setting a desired threshold and trial for different weighting factors until PAPR drops below threshold. However, in this technique there is potential problem with decoding the signal in the presence of noise. In this technique the signal is independent, distortion-less, and does not require any complex optimization techniques. Disadvantages are SLM requires V+1(including un-modified data block) number of IFFTs, and [log₂V] side information is required for each data block, Complexity increased by increasing V [25, 26, and 27]. SLM is more effective than the PTS with the same amount of side information.

## 4.5 Coding

To select code words, which reduce PAPR in MC CDMA of the transmitted signal, using SLM scheme extra side information is required. Using Error control coding with SLM without side information PAPR of MCM signal can be reduced. This scheme requires storing best codes, need large look-up tables for encoding and decoding, especially for large number of subcarriers. This scheme also requires exhaustive algorithm for searching to find best codes from the look-up table. This scheme suffers from extensive calculations to find good codes and offsets. Huffman coding is used to reduce PAPR of OFDM transmitted signal for distortion-less scrambling technique by sending encoding table for accurate decoding at the receiver without reducing throughput. Block coding scheme to reduce the PAPR has many advantages such as no side information at the receiver, codes include error control capacity which can reduce BER at the receiver, reduce PAPR and provide good error correction capability having higher code rates [28, 29].

Park et al have proposed using Walsh Hadamard (WH) code for PAPR reduction in MCM. The proposed Hadamard transform scheme may reduce the occurrence of the high peaks when compared with Original MCM signal. Using Hadamard transform reduces PAPR. In addition, it requires no side information at the receiver for MCM system (OFDM). Imran Baig [12] proposed PAPR reduction in OFDM using Zadeoff Chu-matrix transform based pre/post- code techniques which do not require any power increment, complex optimization and side information to be sent for the receiver.

Coding techniques can be used for signal scrambling, such as Golay complementary sequences [30], Shopire-Rudin sequences, and barker codes. They can be used efficiently to reduce the PAPR. However, with the increase in number of carriers, overhead is increased. More practical solutions of the signal scrambling techniques are block coding, selected mapping and PTS. Signal scrambling technique with side information reduces the throughput, since they introduce redundancy, more degradation and more vulnerable to errors. Error control selective mapping scheme (ECSLM) is more effective and avoids the need of side information. In ECSLM as number of control bits increase, the PAPR reduction increases, and complexity increases.

## 4.6 Precoding Transform Techniques

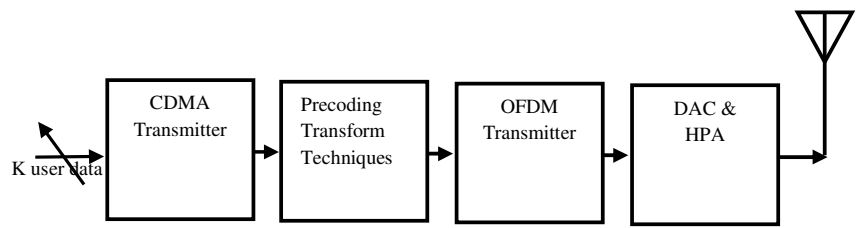

Figure 7 MC CDMA with precoding Transform

Precoding Transform Techniques **(PTT)** are used to reduce PAPR in MC CDMA and OFDM transmitted signal. Using PTT as shown in figure 7, the data is transformed into a new modified data for reduction of PAPR in MC CDMA transmitted signal. The CDMA transmitted signal is fed to precoded transform then followed by IFFT of the OFDM transmitter to reduce PAPR of the MC CDMA transmitted signal. Some of the important Precoded transforms techniques are DCT, DWT and DHT.

### 4.6.1 Discrete Cosine Transform

The data is transformed by a DCT into a new modified data. DCT reduces PAPR of MC CDMA signal without increasing BER of system. DCT de-correlates the data sequence to reduce PAPR in MCM signal. DCT is applied to reduce the autocorrelation of the input sequence before the IFFT operation. Computational complexity is increased by using DCT transform. DCT required $N/2*\log_2N$ multiplications and $3N/2*\log_2(N-N+1)$ additions for N length sequence. DCT does not influence BER [18].

### 4.6.2 Discrete Wavelet Transform

Wavelet transform has a property that it always concentrates energies on parts of signal with sharp changing rates. The DWT transmits only a few numbers of large coefficients which dominates the representation. By using this property the PAPR in Wavelet Multicarrier Modulation (WMCM) system is reduced with little reconstruction loss. Using DWT minimizes ISI and ICI.DWT based MCM can support much higher spectrum efficiency than Fourier based MCM [31].

Haixia Zhang et. al, were the first to report about PAPR reduction in WMCM systems [32, 33] in the open literature. They investigated PAPR reduction methods in MCM system with different orthogonal bases, and the results show that Fourier based MCM outperforms than different orthogonal wavelets based MCM systems. They proposed Wavelet based MCM with clipping to reduce PAPR in WMCM system. The distortion caused by the threshold is measured in terms of Mean Square Error (MSE). B.Sarala et.al [16] have used DWT with companding for MC CDMA system to substantially reduce PAPR There was no need to send side information to the receiver, no distortion, no-loss of data rate. However a little amount of complexity is increased at the receiver.

### 4.6.2 Discrete Hartley Transform

The DHT is linear transform used to reduce the PAPR of the OFDM signal. The DHT transformed CDMA transmitted signal into N real numbers according to N-point DHT. Precoding matrix size is NxN.DHT precoded MCM system shows better PAPR gain when compared with the MCM original system, as well as SLM MCM [34].

## 5. SPREADING SEQUENCE SELECTION CRITERIA

MC CDMA using orthogonal coding without SLM scheme is used to reduce PAPR. Orthogonal spreading codes such as Walsh Hadamard codes, Pseudo Noise (PN) codes, Gold codes, and Golay codes, complementary Golay codes, and PN interferometry code sequences are used in MC CDMA systems [35, 36]. However, these codes are designed to be either orthogonal or otherwise supporting N users with degraded performance. Balasubramanium, Natarajan and Carl R. Nasser [37] proposed by Carrier Interferometry (CI) codes for MC CDMA systems. The CI codes provide flexibility in design as well as increased capacity and strong for MC CDMA wireless networks.MC CDMA with CI codes reduced PAPR by increasing number of users and dynamic range.

In MC CDMA spreading codes influence on the reduction of PAPR of the transmitted signal. Golay spreading codes achieve low PAPR for single user environment. Walsh Hadamard sequences can cause extremely high PAPR values in low traffic downlink MC CDMA system [38]. Lin Yang et.al proposed optimized spreading codes reallocation technique to achieve stable performance for both lightly and fully loaded systems. Walsh Hadamard spreading codes offer better performance on multi user environment. The aim of the spreading sequences is to reduce PAPR for the down link MC CDMA with high data rate transmission, and influence minimization of Multiple Access Interference (MAI) and nonlinear distortion. Low complexity PAPR is required for MC CDMA. Spreading sequences are selected in order to minimize MAI, non-linearity in MC CDMA, and reduce PAPR or crest factor of the transmitted signal [39]. Generally, wireless devices consider two links up-link (Mobile station to Base station) and down-link (Base station to Mobile station), and two sequences orthogonal sequence or non-orthogonal sequence.

## 5.1 Non-orthogonal sequence

### 5.1.1 Gold code sequence

The Gold code sequences are constructed from a preferred pair of m sequence of length $L = 2^n - 1$ by adding modulo-2 (xor) of two m-sequences [40].

G (A, B) = (A, B, $V_0$, $V_1$,-,-, $V_{L-1}$)

With $V_J$ = (A xor $T^j$ B)

And where A = ($a_0$, $a_1$,--, $a_{L-1}$) and

B = ($b_0$, ---, $b_{L-1}$) are

Preferred pair of m- sequence of length L, L+2 Gold sequence of length is available.

Gold codes have three correlation values {-1, -t (n), t (n)-2} where

t(n) = $2^{(n+1)/2}$+1 for n odd.

$2^{(n+2)/2}$+1 for n even.

### 5.1.2 Zadoff-Chu codes

Zadoff-Chu sequences are class of polyphase sequences having optimum correlation properties. Zadoff-Chu sequences of length L have an ideal autocorrelation and constant magnitude √L.

The Zadoff-Chu sequences of length L defined as

Z (k) = $e^{\frac{j2\pi r}{L}\left(\frac{k^2}{2}+qk\right)}$ for L is even

$e^{\frac{j2\pi r(k(k+1)+2qk)}{L}\cdot\frac{1}{2}}$ for L is odd

Where q is integer, k = 0, 1, --, L-1, and r = code index. If L is a prime number integer, the set of Zadoff-Chu is composed of L-1 sequences [12].

## 5.2   Orthogonal Sequence

### 5.2.1 Walsh Hadamard sequences

WH sequence is a set of orthogonal set. Walsh functions are generated using a Hadamard matrix, starting with H1 = [0], The LxL WH matrix is represented as

$$HL = \begin{vmatrix} HL/2 & HL/2 \\ HL/2 & HL/2 \end{vmatrix}$$

WH sequence proposed for synchronous MC CDMA systems.

### 5.2.3 Carrier Interferometry Codes

CI codes provide flexibility in design as well as increased capacity and strong for MC CDMA wireless networks. MC CDMA with CI codes reduces PAPR by increasing number of users in downlink.

$C_k(t)$ is $K^{th}$'s users spreading code defined as

$$C_k(t) = \sum_{i=0}^{N-1} \beta_k^i e^{j2\pi i \Delta f t}$$

Here $\Delta f$ is related that the carriers are orthogonal, i.e., $\Delta f = 1/Tb$; and $\{\beta_k^i, i = 0, 1, ---, N-1\}$ is the kth users spreading code (length N) [37].

### 5.2.4 Orthogonal Gold Codes

The orthogonal codes are developed from a set of original Gold code sequences {1, -1}. Gold code sequence represented as OG (A, B) = (U, $V_0$, $V_1$, ---, $V_{L-2}$) with U = (A, 1).

$V_j$ = (A xor $T^j$ B, 1) where A = ($a_0$, ---, $a_{L-1}$) and B = ($b_0$, ---, $b_{L-1}$) are a preferred pair of m-sequence of length L-1. $T^j$ B is the sequence B after j-chip cyclic shift and xor is modulo-2 addition operator.

### 5.2.5 Pseudo Noise Codes

A m-sequence Pseudo Noise codes (PN), can be generated by using a linear feedback shift register of length n that cycles through all possible $2^n$-1 states , the m-sequence has excellent autocorrelation properties, quasi-orthogonal and periodic [41, 42].

Luis A.Paredes et.al proposed [43] User Reservation approach for PAPR reduction in MC CDMA system for down-link. This method reduces PAPR by 4.5dB (CCDF at $10^{-3}$), and uses Second-order Cone Programming (SOCP) problem, which is computationally costly. User Reservation (UR) approach for PAPR reduction is based on "borrowing" some of the spreading codes of the set of inactive or ideal users for the transmission, so that an adequate linear combination of these codes are added to the active users prior to the IFFT operation. The new signals are selected so that they are orthogonal to the original signal and therefore, it can be removed at the receiver without side information. UR approach reduces the maximum number of users.

## 6. CRITERIA FOR SELECTION OF PAPR REDUCTION TECHNIQUES

In general, reducing the PAPR is always done by the reduction of the transmitted signals, which increase BER at the receiver, and reduce data rate for chosen PAPR reduction scheme. There are several issues that should be considered before the selection of specific PAPR reduction method based on the following technical issues: PAPR reduction capability, Power increase of the transmit signal, BER increase at the receiver, Data loss rate, Computational complexity and other effects.

PAPR reduction capability is very important factor in choosing a PAPR reduction technique. For example, the clipping technique clearly removes the time domain signal peaks, but results in in-band distortion and OBR. Using Tone Reservation technique, power increases in transmit signal after using PAPR reduction. Some techniques may increase BER, and loss in data rate at the receiver if the transmit signal power is fixed. For example in SLM, PTS, and selective scrambling and interleaving techniques the entire data block is lost if the side information is received erroneously. Generally, more complex techniques have better PAPR reduction capability. Depending on the design requirements suitable PAPR reduction technique is chosen.

## 7. COMPARISIONS OF PAPR TECHNIQUES

| PAPR Scheme | PAPR Reduction in dB | Reference Paper serial number |
|---|---|---|
| Amplitude clipping | 2.0 to 3.0 | [5] |
| Clipping & Filtering | 3.9 | [5] |
| Companding (MC CDMA) | 3.5 to 6.5 for μ =2, 3, 5. | [16] |
| SLM(OFDM) | 3.5 | [13] |
| PTS(OFDM) | 4.3 | [13] |
| DCT | 1.0 | [16] |
| DWT | 2.0 | [16] |
| DHT with optimum clipping(OFDM) | 2.7 | [34] |
| Block Coding | 3.7 | [28] |
| Huffman coding(OFDM) | 6.0 | [17] |
| Pulse shaping with RRC(OFDM) | 1.5 | [15] |
| UR with iterative clipping(MC CDMA) | 3.0(10), 5.0 (32), 6.0 (UR optimal) at low load | [43] |
| ECSLM (MC CDMA) | 5.5 | [3] |
| MC CDMA with DCT and companding | 4.6, 5.8, 6.8 dB for μ =2, 3, 5. | [16] |
| MC CDMA with DWT and companding | 5.6, 6.8 dB and 8.1 dB for μ =2, 3, 5. | [16] |

TABLE 1.

The table 1 shows the amount of the PAPR reduction in dB for different PAPR schemes.

## 8. CONCLUSIONS

MC CDMA is a very attractive system for LTE, broadband wireless networks and personal communication beyond 4G. This paper deals with PAPR reduction techniques for Multicarrier modulation transmission systems. Most of them have the potential to reduce PAPR but, at the cost of loss in data rate, increase in transmit signal power, BER, and computational complexity. PAPR of MC CDMA signal can be substantially high. The PAPR can be reduced by using clipping and filtering, and companding which introduces IBR. PTS, SLM and interleaving techniques reduce PAPR with side information and an increased complexity at the receiver and with data loss [13]. Using PTT PAPR is reduced substantially by increasing a little amount of the complexity at the receiver. The PAPR technique of MC CDMA signal transmission should be chosen judiciously. The critical parameters to be optimized are to improve BER and power spectral enhancement with low cost, low complexity, and save bandwidth without losing data.

Authors

**Sarala Beeram** has received her B.Tech. & M.Tech. (Digital Systems & Computer Electronics) from Jawaharlal Technological University, Hyderabad in 1993 and 1998 respectively. She is presently working as an Associate Professor in the Department of ECE in M V S R Engineering College, Hyderabad. Her areas of research include CDMA and Multi Carrier CDMA technologies & Wireless Communications. She has presented more than 11 papers in various national & international conferences. She has also published three papers in international journals.

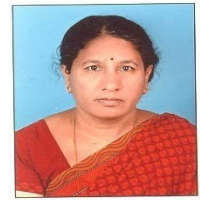

**D.S.Venkateswarlu** received his B.E from Andhra University in 1960, Andhra Pradesh, and M.E from Indian Institute of Science, Bangalore, in 1962, PhD from University of Southampton, UK, in 1967. He has about 13 years industrial experience and about 38 years academic and R&D experience. He has published about 60 papers in peer reviewed journals (National and International) and presented papers in various National and International conferences.

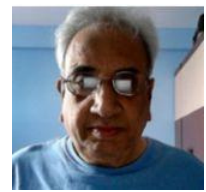

B.N Bhandari his B.Tech from JNTUH, and M.E from Osmania University, PhD from IIT Kargpur. Working as an Associate professor in ECE Dept. and additional controller of examinations, JNTUH. He published several international and national journals, and conference papers

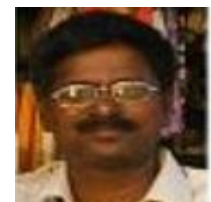